\documentclass[useAMS,usenatbib,usegraphicx]{mn2e}
\usepackage{epsf}

\newcommand{\simgt}{\lower.5ex\hbox{$\; \buildrel > \over \sim \;$}}
\newcommand{\simlt}{\lower.5ex\hbox{$\; \buildrel < \over \sim \;$}}

\title[Cluster finding algorithm]
{A cluster finding algorithm based on the multiband identification of
red sequence galaxies} 

\author[M.~Oguri]
{Masamune Oguri$^{1,2,3}$\thanks{E-mail: masamune.oguri@ipmu.jp} \\
$^1$Research Center for the Early Universe, University of Tokyo, 7-3-1
  Hongo, Bunkyo-ku, Tokyo 113-0033, Japan\\
$^2$Department of Physics, University of Tokyo, 7-3-1 Hongo,
Bunkyo-ku, Tokyo 113-0033, Japan\\
$^3$Kavli Institute for the Physics and Mathematics of the Universe
(Kavli IPMU, WPI), University of Tokyo, Chiba 277-8583, Japan\\
}

\begin{document}

\date{\today}

\voffset- .5in

\pagerange{\pageref{firstpage}--\pageref{lastpage}} \pubyear{}

\maketitle

\label{firstpage}

\begin{abstract}
We present a new algorithm, CAMIRA, to identify clusters of galaxies in
wide-field imaging survey data. We base our algorithm on the stellar
population synthesis model to predict colours of red-sequence galaxies
at a given redshift for an arbitrary set of bandpass filters, with
additional calibration using a sample of spectroscopic galaxies to
improve the accuracy of the model prediction. We run the algorithm
on $\sim 11960$~deg$^2$ of imaging data from the Sloan Digital Sky
Survey (SDSS) Data Release 8 to construct a catalogue of 71743
clusters in the redshift range $0.1<z<0.6$ with richness after
correcting for the incompleteness of the richness estimate greater
than 20. We cross-match the cluster catalogue with external cluster
catalogues to find that our photometric cluster redshift estimates are
accurate with low bias and scatter, and that the corrected richness
correlates well with X-ray luminosities and temperatures. We use the
publicly available Canada-France-Hawaii Telescope Lensing Survey
(CFHTLenS) shear catalogue to calibrate the mass-richness relation
from stacked weak lensing analysis. Stacked weak lensing signals are
detected significantly for 8 subsamples of the SDSS clusters divided
by redshift and richness bins, which are then compared with model
predictions including miscentring effects to constrain mean halo
masses of individual bins. We find the richness correlates well 
with the halo mass, such that the corrected richness limit of 20
corresponds to the cluster virial mass limit of about 
$1 \times 10^{14}h^{-1}M_\odot$ for the SDSS DR8 cluster sample. 
\end{abstract}


\section{Introduction}

Clusters of galaxies have been known to be a useful probe of the
Universe. The mass distribution of clusters is mostly determined by
the dynamics of dark matter, which makes it easier to compare with
theoretical predictions based on $N$-body simulations
\citep[e.g.,][]{navarro97,jing02}. Recent extensive gravitational
lensing analyses have convincingly shown that both the radial density
profile \citep{umetsu11,oguri12,coe12,newman13a,okabe13} and the
degree of non-sphericity \citep{oguri10,oguri12} of massive clusters
are in good agreement with expectations based on the standard
$\Lambda$-dominated cold dark matter model, possibly except for the
dark matter distribution at the very centre where baryonic effects
play a significant role \citep{newman13b}. Clusters of galaxies are
also thought to be one of main probes of dark energy in future surveys
\citep[see][for a review]{weinberg13}, particularly given the
well-understood mass distribution.

Clusters of galaxies can be identified in many different wavelengths. 
For instance, massive clusters of galaxies have efficiently been
identified in X-ray images \citep[e.g.,][]{ebeling01,bohringer04},
because the gas in clusters of galaxies is heated by gravitational
infall to emit thermal bremsstrahlung radiation.  The hot gas in 
clusters also scatters cosmic microwave background (CMB) photons to
distort the CMB spectrum at millimeter and submillimeter
wavelengths. This Sunyaev-Zel'dovich (SZ) effect is rapidly becoming an
efficient way to construct a large sample of clusters, especially at
high redshifts \citep[e.g.,][]{reichardt13,hasselfield13}. A
disadvantage of X-ray and SZ cluster surveys is a lack of redshift
information. Hence these cluster samples should be complemented by
optical or near-infrared data for redshift estimates of individual
clusters.  

Thanks to recent developments of wide-field optical surveys, large
catalogues of clusters are being constructed in optical wavelengths. 
While finding clusters in singe band optical imaging data
\citep[e.g.,][]{abell58} is challenging given the low number density
contrast of galaxies, one can identify clusters much more easily and
securely by utilizing multi-band optical data
\citep[e.g.,][]{gladders00,liu08,milkeraitis10,murphy12,jian14}. 
Furthermore, multi-band optical cluster selections usually provide
good photometric redshifts of clusters, and thereby enable to construct
three-dimensional cluster catalogues.

Many applications of these cluster catalogues, including constraints
on cosmological parameters \citep[e.g.,][]{vikhlinin09,mantz10,rozo10},
require knowledge of scaling relations between observables and masses
of clusters. Stacked weak lensing provides a powerful means of
accurate calibration of such scaling relations 
\citep[e.g.,][]{johnston07,leauthaud10,ford14,covone14}. In
particular, we can use the same imaging data for both identifying
clusters and weak lensing mass calibrations in wide-field optical
surveys. In these surveys, by measuring a mean tangential shear
profile around a sample of clusters, we can accurately constrain the
average mass of the cluster sample \citep{sheldon09,oguri11a,rozo11}. 
This can be regarded as another advantage of selecting clusters of
galaxies in optical wavelength. A caveat is that the orientation bias
of optically selected clusters can lead to the overestimation of the
average mass by up to $\sim 5\%$ \citep{dietrich14}. 

In this paper, we present a new optical cluster finding algorithm. The
algorithm, which we name CAMIRA (Cluster finding algorithm based on
Multi-band Identification of Red-sequence gAlaxies), is essentially a
red-sequence method, and has a flexibility to allow to use an
arbitrary set of filters. For this purpose, we base our algorithm on
the stellar population synthesis (SPS) model. The SPS model, after
appropriate calibrations to improve the accuracy, is used to predict
colours and stellar masses of red-sequence galaxies. Then each galaxy
in the image is fitted to the SPS model to compute the likelihood of
being a red-sequence galaxy at a given redshift.  The use of only the
red-sequence galaxies is because it is expected to reduce the scatter
in the mass-richness relation \citep{rozo09,rykoff12}. Our method also
implements an algorithm for finding the brightest cluster galaxy (BCG)
and takes account of masking effects.  Our algorithm is similar to the
recently published redMaPPer method \citep{rykoff14,rozo14} in several
ways, though we note that our algorithm is developed mostly
independently of redMaPPer.  

We apply our method, CAMIRA, to the Sloan Digital Sky Survey
\citep[SDSS;][]{york00} data to construct a cluster catalogue in the
redshift range $0.1<z<0.6$. Specifically we use imaging data from the
SDSS Data Release 8 \citep[DR8;][]{aihara11} which covers more than
10000~deg$^2$ of the sky. There have already been many algorithms
that were applied to the SDSS data to produce large cluster catalogues 
\citep{goto02,miller05,koester07a,koester07b,dong08,wen09,szabo11,wen12,hao09,hao10,rykoff14,rozo14},
suggesting that the SDSS dataset is ideal for developing and testing
new algorithms. Another advantage of the SDSS is the availability of
a large number of spectroscopic measurements of red galaxies, which
are in our algorithm used to calibrate the SPS model. We then use
various X-ray data as well as the public Canada-France-Hawaii
Telescope Lensing Survey \citep[CFHTLenS;][]{heymans12} shear
catalogue to test and characterize our SDSS cluster catalogue.

The outline of this paper is as follows. In Section~\ref{sec:algorithm}, 
we describe our cluster finding algorithm in
detail. Section~\ref{sec:catalogue} presents our cluster catalogue in
the SDSS DR8. Section~\ref{sec:test} describes testing of the
algorithm mostly using X-ray data. We also conduct weak lensing
analysis of the SDSS cluster sample in Section~\ref{sec:weaklens}.
In Section~\ref{sec:summary}, we summarize our results. The SDSS DR8
cluster catalogue is presented in
Appendix~\ref{sec:table}. Throughout the
paper we adopt the standard $\Lambda$-dominated flat cosmological
model with the matter density $\Omega_M=1-\Omega_\Lambda=0.28$, the
dimensionless Hubble constant $h=0.7$, the baryon matter density
$\Omega_b=0.042$, the spectral index $n_s=0.96$, and the normalization
of the matter fluctuation $\sigma_8=0.8$.

\section{Algorithm}
\label{sec:algorithm}

\subsection{Modelling red-sequence galaxies}
\label{sec:rsmodel}

We use the SPS model of \citet{bruzual03} to model the spectral energy
distribution of red-sequence galaxies. The advantage of using the SPS
model is that one can easily compute colours in an arbitrary combination
of filters, which is essential for multi-band selections of
red-sequence galaxies as considered in this paper.  Throughout the
paper we assume the Salpeter initial mass function. 

The SPS model characterizes the properties of galaxies by several
parameters, including the age of the galaxy, star formation history,
metallicity ($Z$), the stellar mass ($M_*$), and the dust extinction. 
Our basic strategy is to adjust these parameters to reproduce the
observed colours of red-sequence galaxies, and use the model for
calculating the likelihood of galaxies being in the red-sequence as a
function of redshift. While complicated models contain more degree of
freedom to calibrate the SPS model to reproduce observed red-sequence
colours, here we adopt a rather simple model with a single
instantaneous burst at the formation redshift $z=z_f$ and no dust
extinction, as the model appears to be already good enough to model
the red-sequence (see below).

The colour-magnitude diagram of red-sequence galaxies is known to
exhibit the so-called ``tilt'', i.e., the galaxy colours changes
slightly as a function of magnitude, which originates from the mass
dependence of metallicity \citep{kodama97,stanford98}. We include the
tilt by assuming the following functional form for metallicity:  
\begin{equation}
\log Z_{\rm SPS} = \log Z_{11}+
a_Z\left[\log (M_{*,{\rm in}}/10^{11}M_\odot)\right],
\end{equation}
where $M_{*,{\rm in}}$ is the input stellar mass, or put another way, the
total stellar mass formed at $z_f$, and $Z_{11}$ specifies the
normalization of metallicity, i.e., $Z_{11}$ is metallicity of
galaxies with $M_{*,{\rm in}}=10^{11}M_\odot$. The input stellar mass
$M_{*,{\rm in}}$ in general differs from the stellar mass at the age
considered, $M_*$, because a fraction of the total stellar mass
originally formed is converted to gas as a consequence of stellar
evolution. For technical reasons, throughout the paper we use
$M_{*,{\rm in}}$ rather than $M_*$ as a model parameter.

We determine the model parameters $z_f$, $\log Z_{11}$, $a_Z$, by examining
colour-magnitude diagrams in several massive clusters. Specifically,
we choose $z_f=3$, $\log Z_{11}=-2$, and $a_z=0.15$, which are found
to reproduce observed colour-magnitude relations of cluster member
galaxies in the SDSS data reasonably well. 

It is known that the colour-magnitude relation involves an intrinsic
scatter. We model the intrinsic scatter by the scatter of the
metallicity. Again, based on the examination of colour-magnitude
relations for SDSS clusters, we adopt the scatter of $\sigma_{\log
  Z}=0.14$ to model the intrinsic scatter. We also restrict the
stellar mass range when fitting, 
$M_{*,{\rm min}}<M_{*,{\rm in}}<M_{*,{\rm max}}$. Here we set $\log
(M_{*,{\rm max}}/M_\odot)=13.5$ and 
$\log (M_{*,{\rm min}}/M_\odot)=9.5$, which cover the stellar mass
filter range introduced below. 

\subsection{Calibrating colours}
\label{sec:modelcalib}

The SPS model predicts red-sequence galaxy colours reasonably well,
but is never perfect. Therefore it is essential to calibrate galaxy
colours using observed colours of galaxies with spectroscopic
redshifts. 

In this paper, we quantify the likelihood of each galaxy being
red-sequence galaxies at redshift $z$ by the following chi-square
\begin{equation}
  \chi^2=\sum_{i=1}^{N_{\rm fil}}\frac{(m_{i,{\rm obs}}-m_{i,{\rm
        SPS}}-\delta m_{i,{\rm resi}})^2}{\sigma_{m_{i,{\rm
          obs}}}^2+\sigma_{m_{i,{\rm resi}}}^2}+\frac{(\log Z_{11}-\log
    \bar{Z}_{11})^2}{\sigma_{\log Z}^2},
\label{eq:chi2}
\end{equation}
where $i$ runs over photometric bands of the galaxy catalogue, $N_{\rm
  fil}$ is the total number of photometric bands, $m_{i,{\rm obs}}$ and 
$\sigma_{m_{i,{\rm obs}}}$ are observed magnitude and its error in the $i$-th
  band, $m_{i,{\rm SPS}}$ is the SPS model prediction at redshift $z$,
 and $\log \bar{Z}_{11}=-2$ and $\sigma_{\log Z}=0.14$ (see
 Section~\ref{sec:rsmodel} for details). In addition, $\delta m_{i,{\rm
   resi}}$ and $\sigma_{i,{\rm resi}}$ are included to account for the
 imperfectness of the SPS model. 

Our SPS model is calibrated by estimating $\delta m_{\rm resi}$ and 
$\sigma_{\rm resi}$ as a function of rest-frame wavelength and
redshift. For each spectroscopic galaxy, we minimize $\chi^2$ by
varying $Z_{11}$ and $M_{*,{\rm in}}$. We then fit residuals of
magnitudes for a sample of spectroscopic galaxies as a function of 
rest-frame wavelength $\lambda$ and redshift $z$. Specifically,
we divide the sample into different redshift bins, and in each
redshift bin $z_j$ we fit the residuals to polynomials in $\lambda$:
\begin{equation}
  \delta m_{\rm resi,fit}(\lambda,
  z_j)=\sum_{i=1}^{n_f}a_i(z_j)
\left(\lambda-\lambda_0\right)^i,
\label{eq:poly}
\end{equation}
with $a_i(z_j)$ being polynomial coefficients. Throughout the paper we
fix $\lambda_0=5000$~{\AA}. We then construct smooth functions of
polynomial coefficients $a_i(z)$ as a function of redshift by the
spline interpolation. 

The scatter $\sigma_{\rm resi}$ describes the scatter of spectral energy 
distributions of red-sequence galaxies that is unaccounted in our SPS
model. We divide magnitude residuals of the spectroscopic galaxies
into rest-wavelength bins and compute a scatter in each bin with
a weight of $1/(\sigma_{\rm obs}^2+\sigma_{\rm resi}^2)$ for each
residual. In this procedure we also remove $2.5\sigma$ outliers. In
each bin, this calculation is performed iteratively until the value
of $\sigma_{\rm resi}$ converges. The bin size of
$\Delta\lambda=400$~{\AA} is adopted.  

In order to minimize the effect of outliers, such as spectroscopic
galaxies that are outside the red-sequence, we iteratively compute
$\delta m_{\rm resi}$ and $\sigma_{\rm resi}$. In the first round, we
compute $\chi^2$ with  $\delta m_{\rm resi}=\sigma_{\rm resi}=0$, but
only include galaxies with best-fit metallicity of $-1.65<\log
Z_{11}<-2.35$, corresponding to $2.5\sigma$ in the metallicity
scatter. We then repeat the calculation of $\chi^2$ including
residuals and scatters estimated in the previous pass and refine 
these by using galaxies with $\chi^2<\chi^2_{\rm max,resi}$ for
residual fitting and $\chi^2<\chi^2_{\rm max,\sigma}$ for estimating
their scatter, with $\chi^2_{\rm max, resi}=4$ and $\chi^2_{\rm
  max,\sigma}=20$ throughout this paper. The second pass is repeated
twice to further refine the residual estimate.  

Equation~(\ref{eq:chi2}) does not include the off-diagonal element of
the covariance matrix of model magnitude errors. In practice, we
expect some correlated model errors between different bands, but we
assume that those correlated errors are taken care of by including
metallicity in the model fitting, because shifting metallicity
systematically changes colours of red-sequence galaxies. Put another
way, our working assumption is that correlated errors of magnitudes of
red-sequence galaxies between different bands can be modelled by the
scatter of metallicity which is included in our fitting
procedure. Indeed, we check residual distributions of spectroscopic
galaxies for the calibration and find no significant correlations
between residuals of different bands, which supports our working
assumption. 

\subsection{Constructing a richness map}

\begin{figure}
\begin{center}
 \includegraphics[width=0.95\hsize]{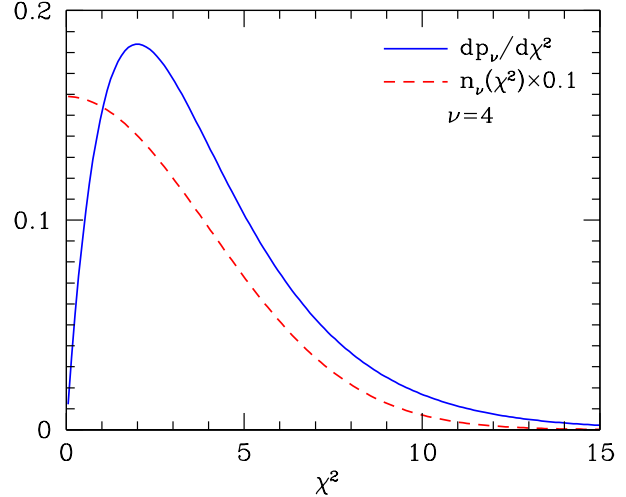}
\end{center}
\caption{The $\chi^2$ distribution $dp_\nu/d\chi^2$
  (equation~\ref{eq:chi2dist}; {\it solid line}) and the number parameter
  $n_\nu(\chi^2)$ (equation~\ref{eq:chi2num}; {\it dashed line}) as a
  function of $\chi^2$. The number parameter $n_\nu(\chi^2)$ is
  multiplied by $0.1$ for illustrative purpose. The degree of freedom
  of $\nu=4$ is assumed. 
  \label{fig:c2func}}
\end{figure}

For red-sequence galaxies, $\chi^2$ computed by
equation~(\ref{eq:chi2}) at the galaxy redshifts should obey the
$\chi^2$ distribution with  $\nu=N_{\rm fil}-1$ degrees of freedom:
\begin{equation}
\frac{dp_\nu}{d\chi^2}=\frac{1}{2^{\nu/2}\Gamma(\nu/2)}
e^{-\chi^2/2}(\chi^2)^{\nu/2-1},
\label{eq:chi2dist}
\end{equation}
where $\Gamma(x)$ denotes the Gamma function. Bearing this in mind, 
we define cluster member galaxy ``number'' parameter $n_\nu(\chi^2)$
as 
\begin{equation}
n_\nu(\chi^2)=\frac{2^{3\nu/4}}{\nu^{\nu/2}U(\nu/4,1/2,\nu^2/8)}
e^{-(\chi^2)^2/2},
\label{eq:chi2num}
\end{equation}
with $U(a,b,x)$ being the confluent hypergeometric function of the
second kind. The normalization of $n_\nu(\chi^2)$ is chosen so as to
satisfy
\begin{equation}
\int_0^\infty n_\nu(\chi^2)\frac{dp_\nu}{d\chi^2}d\chi^2=1.
\end{equation}
Thus when we sum up $n_\nu(\chi^2)$ over $N_{\rm mem}$ cluster member
galaxies we expect to have 
\begin{equation}
\left\langle \sum_{i=1}^{N_{\rm mem}} n_\nu(\chi^2)\right\rangle =
N_{\rm mem}\int n_\nu(\chi^2)\frac{dp_\nu}{d\chi^2} d\chi^2 = N_{\rm mem}.
\end{equation}
We show $dp_\nu/d\chi^2$ and $n_\nu(\chi^2)$ in
Figure~\ref{fig:c2func}. In reality galaxy catalogues contain
non-member galaxies. The contribution of these non-members is
negligible if they have large enough $\chi^2$, i.e., $\chi^2\gg \nu$,
but in practice foreground and background galaxies can make
non-negligible contributions to the sum of $n_\nu(\chi^2)$.  
We account for this by subtracting the background level, as
we will describe in more detail below. 

We count the number of member galaxies in a specific stellar mass
range. The lower mass limit should correspond to $\sim 0.2 L_*$
because it was shown to be optimal in terms of richness measurements
\citep{rykoff12}. The upper mass limit is also important to reduce
possible projection effects in selecting member galaxies. Thus we
choose our stellar mass filter as
\begin{equation}
F_M(M_{*,{\rm in}})=\exp\left[-\left(\frac{M_{*,{\rm in}}}{M_{\rm
        h}}\right)^4
-\left(\frac{M_{\rm l}}{M_{*,{\rm in}}}\right)^4\right],
\label{eq:filter_sm}
\end{equation}
where $M_{\rm h}=10^{13}M_\odot$ and $M_{\rm l}=10^{10.2}M_\odot$ are
adopted in this paper. 

The number of member galaxies should be counted within some aperture
that roughly corresponds to radii of clusters of our interest. On the
other hand, the background level must be subtracted to correctly
estimate richness. The background level, however, is not uniform but has
large-scale structure. A compensated filter estimates the background
level in an annulus just outside the aperture, and hence properly
takes account of the non-uniformity of the background. In this paper,
we adopt  spatial filter of the following functional form:
\begin{equation}
F_R(R)=\frac{\Gamma\left[n/2,(R/R_0)^2\right]
-(R/R_0)^ne^{-(R/R_0)^2}}{\Gamma\left(n/2,0\right)}.
\label{eq:filter}
\end{equation}
The filter is normalized as $F_R(0)=1$. The filter 
$F_R(R)$ with different values of $n$ are plotted in
Figure~\ref{eq:filter}.  Throughout the paper we adopt $n=4$ and a
fixed scale radius $R_0=0.8h^{-1}{\rm Mpc}$ in physical unit. 
While a possible extension of our algorithm is to include the richness
dependence of the scale radius \citep[e.g.,][]{rykoff12}, here we
adopt the fixed scale radius for simplicity.

\begin{figure}
\begin{center}
 \includegraphics[width=0.95\hsize]{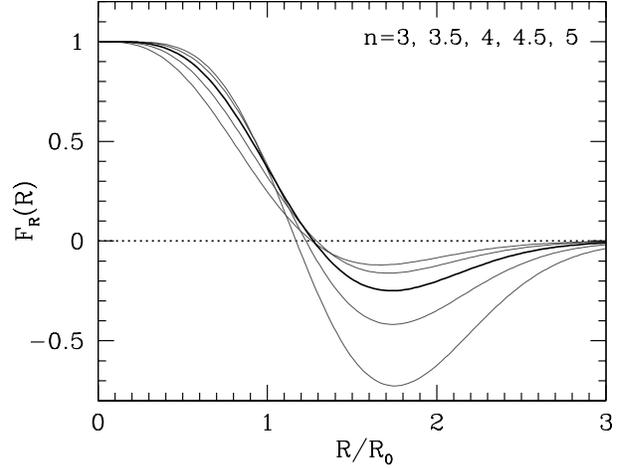}
\end{center}
\caption{The compensated filter $F_R(R)$
  (equation~\ref{eq:filter}) used to derive the richness map. We show
  filter functions with the parameter $n=3$, $3.5$, $4$, $4.5$, and
  $5$, where $n=5$ curve has the most prominent negative peak. The
  filter function for our fiducial choice of $n=4$ is shown by the
  thick solid line. 
  \label{fig:file}}
\end{figure}

Now we can construct a richness map by adding up the number parameter
with a weight of these filter functions
\begin{equation}
N_{\rm mem}(\bmath{\theta}, z)=\sum_i n_\nu(\chi^2_i; \bmath{\theta}_i, z)
F_M(M_{*,i})
F_R(D_A|\bmath{\theta}_i-\bmath{\theta}|),
\end{equation}
where $n_\nu(\chi^2_i; \bmath{\theta}_i, z)$ is the number parameter 
(equation~\ref{eq:chi2num}) for a galaxy at $\bmath{\theta}_i$
evaluated at redshift $z$ and $D_A=D_A(z)$ is an angular diameter
distance at redshift $z$. The parameters $\chi^2_i$ and $M_{*,i}$ are
best-fit $\chi^2$ and $M_{*{\rm in}}$ for the $i$-th galaxy at
redshift $z$. In practice, for each redshift slice we compute the
two-dimensional richness map by creating a number density map in a
regular grid and then use the Fast Fourier Transform to obtain 
$N_{\rm mem}(\bmath{\theta}, z)$.  

\subsection{Masking}
\label{sec:masking}

In a real survey there are many patches of the sky in the survey region
where the catalogue contains no galaxy. These masked regions originates
from a combination of various effects, such as the existence of bright
stars, gaps between pointings or CCDs, and observing data taken in bad
weather conditions. Since clusters of galaxies are extended on the
sky, estimating the impact of these mask regions on our cluster
finding algorithm is not obvious. In this Section we describe our
algorithm to account for the mask regions.

We create a mask map using the input galaxy catalogue. At each point on
the sky we compute the total number of galaxies in the catalogue
within a circle of the radius $\theta_{\rm mask}$, and mask that point
if there is no galaxy within $\theta_{\rm mask}$. In this paper we fix
$\theta_{\rm mask}=1'$. From this procedure we construct a mask map 
$S(\bmath{\theta})$ with $S(\bmath{\theta})=1$ and $0$ denoting
unmasked and masked regions, respectively. We then correct for the
masking effect in the estimates of the richness. The masking-corrected
richness $ \hat{N}_{\rm mem}$ is computed as 
\begin{eqnarray}
\hat{N}_{\rm mem}(\bmath{\theta}, z)&=&\sum_i \frac{1}{f_{\rm mask}} 
n_\nu(\chi^2_i; \bmath{\theta}_i, z)
F_M(M_{*,i})\nonumber\\
&&\times F_R(D_A|\bmath{\theta}_i-\bmath{\theta}|),
\label{eq:nmem}
\end{eqnarray}
where
\begin{equation}
 f_{\rm mask}=
  \left\{
    \begin{array}{ll}
    f_{\rm mask, c} & (F_R(D_A|\bmath{\theta}_i-\bmath{\theta}|)>0), \\
    f_{\rm mask, b} & (F_R(D_A|\bmath{\theta}_i-\bmath{\theta}|)<0),
  \end{array} \right.
\end{equation}
\begin{equation}
  f_{\rm mask,c}=\frac{\int_{F_R>0} d\bmath{\theta}'
S(\bmath{\theta}')F_R(D_A|\bmath{\theta}'-\bmath{\theta}|)}
{\int_{F_R>0} d\bmath{\theta}'F_R(D_A|\bmath{\theta}'
-\bmath{\theta}|)},
\end{equation}
\begin{equation}
  f_{\rm mask,b}=\frac{\int_{F_R<0} d\bmath{\theta}'
S(\bmath{\theta}')F_R(D_A|\bmath{\theta}'-\bmath{\theta}|)}
{\int_{F_R<0} d\bmath{\theta}'F_R(D_A|\bmath{\theta}'
-\bmath{\theta}|)}.
\end{equation}
When the masking area is too large, $f_{\rm mask}\ll 1$, the richness
estimate becomes highly uncertain. Thus we impose minimum values on
both $f_{\rm mask,c}$ and $f_{\rm mask,b}$ below which the richness map
is masked. We adopt the minimum values of 0.6 and 0.2 for $f_{\rm
  mask,c}$ and $f_{\rm mask,b}$, respectively. 

\subsection{Refining cluster candidates}
\label{sec:refine}

We identify cluster candidates from peaks in the three-dimensional
richness map, $\hat{N}_{\rm mem}(\bmath{\theta}, z)$. We then refine
redshift and richness estimates of each peak as follows. 

First we sort the cluster candidate list in descending order of the
peak richness. For each peak at $\bmath{\theta}=\bmath{\theta}_{\rm
  p}$, we start with refining the cluster redshift estimate. Following
\citet{rykoff14}, we obtain a new cluster redshift by maximizing the
following likelihood
\begin{equation}
\ln \mathcal{L}_z=-\frac{1}{2}\sum_{i} w_i\chi^2_i(\bmath{\theta}_i,
z)\Theta\left[F_R(D_A|\bmath{\theta}_i-\bmath{\theta}_{\rm p}|)
  \right],
\label{eq:zcl_like}
\end{equation}
where the summation runs over galaxies (each located at
$\bmath{\theta}_i$). The weight $w_i$ is introduced so that we only
use high significance cluster member galaxies for estimating the
cluster redshift. It is defined as
\begin{equation}
w_i=\frac{1}{1+\exp\left\{(n_{\rm th}-n_{{\rm mem},i})/\sigma_n\right\}},
\end{equation}
where $n_{{\rm mem},i}$ is the number parameter for each galaxy
\begin{equation}
n_{{\rm mem},i}=
\frac{1}{f_{\rm mask}} n_\nu(\chi^2_i; \bmath{\theta}_i, z)
F_M(M_{*,i})
F_R(D_A|\bmath{\theta}_i-\bmath{\theta}_{\rm p}|),
\end{equation}
and $n_{\rm th}$ is defined such that
\begin{equation}
\sum_{n_{{\rm mem},i}>n_{\rm th}} n_{{\rm mem},i}=f_n \hat{N}_{\rm
  mem}(\bmath{\theta}_{\rm p}, z).
\end{equation}
We adopt $f_n=0.5$ and $\sigma_n=0.05$. The weigh $w_i$ becomes close
to $1$ and $0$ for large and small number parameters, respectively.
The new cluster redshift $z_{\rm cl}$ is the redshift that maximizes
the likelihood defined by equation~(\ref{eq:zcl_like}).

Next we identify the BCG of the cluster. For each galaxy at
$\bmath{\theta}=\bmath{\theta}_i$ we compute
the likelihood of being the BCG by fixing the redshift to $z_{\rm cl}$
\begin{equation}
\ln \mathcal{L}_{\rm BCG}=-\frac{\left[\log (M_{*,i}/M_{*,{\rm
 BCG}}) \right]^2}{2\sigma_{\log M}^2}+\ln n_\nu(\chi^2_i)
-\frac{R_i^2}{\sigma_R^2},
\label{eq:bcg_like}
\end{equation}
where $\chi^2_i$ and $M_{*,i}$ are best-fit $\chi^2$ and $M_{*,{\rm
    in}}$ for redshift $z=z_{\rm cl}$, respectively, and $R_i=D_A(z_{\rm
  cl})\left|\bmath{\theta}_i-\bmath{\theta}_{\rm p}\right|$ is the
physical distance between the peak and the galaxy. The first term in
the right hand side of equation~(\ref{eq:bcg_like}) aims at selecting
massive (bright) galaxies as a BCG candidate. The second terms simply
indicates the BCG should be a cluster member galaxy at high
significance. The last term is introduced to assure that the position
of the BCG is not too far from the peak of the richness map. The
parameters in these terms should be chosen empirically so as to select
the BCG effectively. In this paper we tentatively assume $\log(M_{*,{\rm
    BCG}}/M_\odot)=12.3$, $\sigma_{\log M}=0.3$, and 
$\sigma_R=0.3h^{-1}{\rm Mpc}$. The likelihood function
(equation~\ref{eq:bcg_like}) and the parameters in have been
determined rather empirically, and are subject to improvements by
careful analysis of cluster centring. Also the so-called ``blue BCGs''
will not be efficiently selected by this algorithm because we impose the
condition that the BCG be a red-sequence galaxy ($\ln
n_\nu(\chi^2_i)$ in equation~\ref{eq:bcg_like}).

Once the BCG candidate is obtained, we again estimate the cluster
redshift via equation~(\ref{eq:zcl_like}) replacing the peak position
$\bmath{\theta}_{\rm p}$ with the candidate BCG position
$\bmath{\theta}_{\rm BCG}$. With the refined cluster redshift $z_{\rm
  cl}$ we again search for the BCG which maximizes the likelihood
defined by equation~(\ref{eq:bcg_like}). This procedure is repeated
until the solution converges. Finally we define the richness of this
cluster by $\hat{N}_{\rm mem}$ (equation~\ref{eq:nmem}) computed at
the BCG position $\bmath{\theta}_{\rm BCG}$ and redshift $z_{\rm cl}$,
i.e., $\hat{N}_{\rm mem}=\hat{N}_{\rm mem}(\bmath{\theta}_{\rm BCG},
z_{\rm cl})$.

After the final cluster candidate is obtained, we percolate the catalogue
to ensure that no cluster is multiply counted. For each galaxy we
assign a weight factor $w_{\rm mem}$ that scales similar to a
membership probability as
\begin{equation}
 w_{\rm mem}=
    n_\nu(\chi^2_i)F_M(M_{*,i})F_R(D_A|\bmath{\theta}_i-\bmath{\theta}_{\rm
      BCG}|), 
\end{equation}
for $F_R>0$, and $w_{\rm mem}=0$ for $F_R<0$. In the examinations of
lower richness peaks the number parameter of these galaxies are
multiplied by an additional factor of $1-\sum w_{\rm mem}$ in order to
avoid double counting of cluster member galaxies. Galaxies with $\sum
w_{\rm mem}\geq 1$ are not used in the subsequent analysis. 

\section{Cluster catalogue in SDSS DR8}
\label{sec:catalogue}

\subsection{Data}

We apply our cluster finding algorithm CAMIRA to imaging data of SDSS
DR8 \citep{aihara11}. The input galaxy catalogue include model
magnitudes ({\tt MODEL\_MAG}) and their errors for SDSS $ugriz$-band. 
We exclude galaxies with any of the following flags; 
{\tt SATURATED}, {\tt SATUR\_CENTER}, {\tt BRIGHT}, and {\tt
  DEBLENDED\_AS\_MOVING}. 
We only use galaxies with extinction-corrected $i$-band magnitude
brighter than 21.0 and its error smaller than 0.2. The dust extinction
map of \citet{schlegel98} is used for the Galactic extinction
correction. We use all galaxies in the RA ranges between $310$ and
$50$~deg and between $110$ and $270$~deg and the Dec range between
$-11$ and $69$~deg, which fully cover the main survey regions of SDSS
both in the North and South Galactic Caps. 

In the SDSS footprint there are several bad regions, such as regions
near bright stars and nearby galaxies, where the photometric
calibration contains some problems. In this paper we do not mask these
regions. Therefore, for some applications of the cluster catalogue,
such as angular clustering measurements, one may have to apply
additional masks to remove these bad regions.   

\subsection{Calibration}
\label{sec:sdsscalib}

\begin{figure}
\begin{center}
 \includegraphics[width=0.95\hsize]{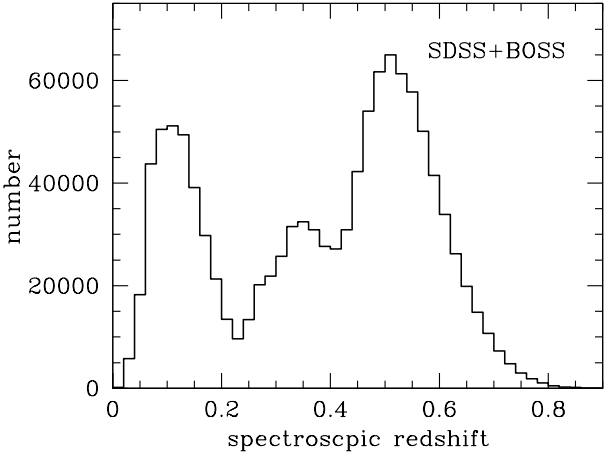}
\end{center}
\caption{The number distribution of SDSS spectroscopic galaxies, which
  are used for the colour calibration, as a function of
  redshift. Three peaks at $z\sim 0.1$, $z\sim 0.35$, and $z\sim 0.5$
  correspond to typical redshifts of SDSS main galaxy sample
  \citep{strauss02}, SDSS luminous red galaxy sample
  \citep{eisenstein01}, and BOSS CMASS galaxy sample \citep{dawson13}, 
  respectively. See Section~\ref{sec:modelcalib} for details.
  \label{fig:zdist_sdss}}
\end{figure}

\begin{figure}
\begin{center}
 \includegraphics[width=0.95\hsize]{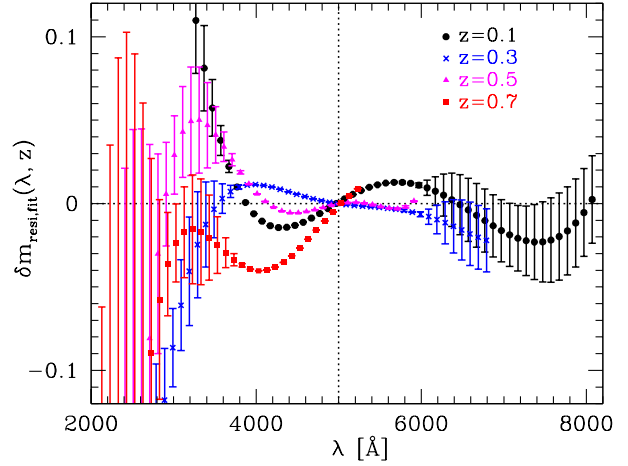}
\end{center}
\caption{Residuals $\delta m_{\rm resi,fit}$ and model scatter
  $\sigma_{\rm resi}$ as a function of rest-frame wavelength
  $\lambda$, which are obtained from the SPS fitting to SDSS
  spectroscopic galaxies (see Figure~\ref{fig:zdist_sdss} for the
  number distribution). Details of the calibration to derive the
  residuals are described in Section~\ref{sec:modelcalib}. We show
  residuals and scatters for 4 different redshifts, $z=0.1$ ({\it
    filled circles}), $0.3$ ({\it crosses}), $0.5$ ({\it filled
    triangles}), and $0.7$ ({\it filled squares}). Residuals for each 
  redshift are shown only in the wavelength range covered by the 
  SDSS $ugriz$-band filters. 
  \label{fig:resi}}
\end{figure}

We use spectroscopic galaxy catalogues from the BOSS DR10
\citep{ahn14} as well as SDSS DR7 \citep{abazajian09} for the
calibration of galaxy colours as described in
Section~\ref{sec:modelcalib}. Since we are interested in red-sequence
galaxies, we apply a rough colour cut  
\begin{equation}
g-r>
  \left\{
    \begin{array}{ll}
    0.6+(5/3)z_g & (z_g<0.3),\\
    1.1 & (z_g>0.3),\\
  \end{array} \right.
\end{equation}
\begin{equation}
g-r< 4.0,
\end{equation}
\begin{equation}
r-i> 0.3,
\end{equation}
\begin{equation}
u-g>1.6 \;\;\;\mbox{for $z_g<0.25$},
\end{equation}
where $z_g$ denotes a spectroscopic redshift of each galaxy.
The calibration process (Section~\ref{sec:modelcalib}) is performed
iteratively to remove contributions from outliers, and hence the
colour cut here is intended to remove only obvious non red-sequence
galaxies. We use 1152403 galaxies after the colour cut to calibrate
galaxy colours, adopting the order of $n_f=5$ for the polynomial 
fitting (equation~\ref{eq:poly}). The redshift bin width is $\Delta
z=0.02$, and the calibration is done for the redshift range of
$0.02<z<0.82$. The redshift distribution of these spectroscopic
galaxies is shown in Figure~\ref{fig:zdist_sdss}.  
We show resulting residuals $\delta m_{\rm resi,fit}$
(equation~\ref{eq:poly}) as well as scatter $\sigma_{\rm resi}$ 
in Figure~\ref{fig:resi}.

\begin{figure}
\begin{center}
 \includegraphics[width=0.95\hsize]{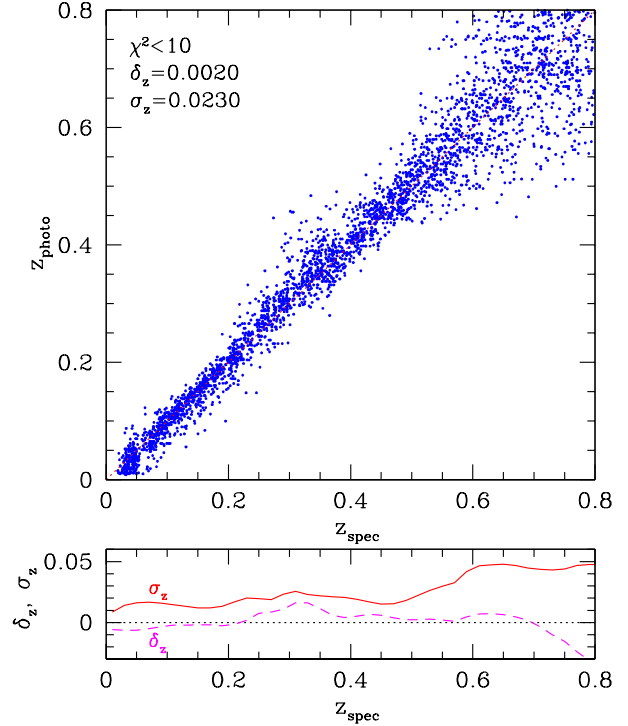}
\end{center}
\caption{Photometric redshifts of SDSS and BOSS galaxies that are
  derived by minimizing $\chi^2$ (equation~\ref{eq:chi2}) are
  compare with their spectroscopic redshifts. The comparison is  
  done only for galaxies with good fits, $\chi^2<10$. Here we show
  results for a subsample of 3641 galaxies which are randomly selected
  to achieve an almost uniform distribution over the redshift. 
  The lower panel shows the mean $\delta_z$ ({\it solid}) and the
  scatter $\sigma_z$ ({\it dashed}) of residuals $(z_{\rm photo}-
  z_{\rm spec})/(1+z_{\rm spec})$ as a function of redshift, computed
  using all the spectroscopic galaxies rather than the subsample. 
  \label{fig:zcomp}}
\end{figure}

As a sanity check, we derive ``photometric redshifts'' of these
spectroscopic galaxies by finding redshifts that minimize $\chi^2$
defined by equation (\ref{eq:chi2}), after the calibration of galaxy
colours as described above, and compare them with their spectroscopic
redshifts. Figure~\ref{fig:zcomp} shows the comparison of the
photometric redshifts $z_{\rm photo}$ with the spectroscopic redshifts
$z_{\rm spec}$ from the SDSS and BOSS. The Figure indicates that our
model, once the calibration is properly done, recovers true redshifts
very well. To quantify the accuracy of the photometric redshifts, we
compute the mean $\delta_z$ and the scatter $\sigma_z$ of residuals
$(z_{\rm photo}- z_{\rm spec})/(1+z_{\rm spec})$, with 3$\sigma$
clipping to exclude the effect of outliers. We find $\delta_z=0.0020$
and $\sigma_z=0.0230$, which are sufficiently good for our purpose. 

\subsection{Richness correction}
\label{sec:richcor}

The magnitude-limited nature of imaging surveys suggests that the
richness estimate can be incomplete particularly at high redshifts.
In the case of SDSS, the smooth stellar mass cut at $M_{*,{\rm in}}\sim
M_l=10^{10.2}M_\odot$ (see equation~\ref{eq:filter_sm}) indicates that the
richness estimate is nearly complete only at $z\la 0.25$. Here we
provide a scheme to empirically correct for the richness
incompleteness as a function of redshift.

\begin{figure}
\begin{center}
 \includegraphics[width=0.95\hsize]{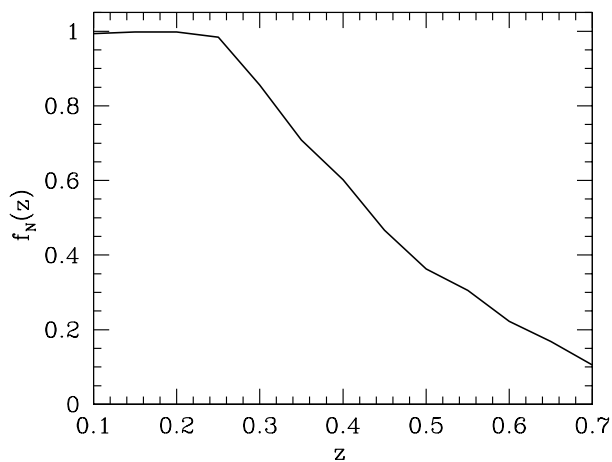}
\end{center}
\caption{The richness correction factor $f_N(z)$
  (equation~\ref{eq:fncor}) as a function of redshift for our sample
  of SDSS clusters.  
  \label{fig:fncor}}
\end{figure}

First for each redshift we derive the stellar mass function
$d\phi/dM_{*,{\rm in}}$ of red-sequence galaxies by summing up the number
parameter $n_\nu(\chi^2)$ (equation~\ref{eq:chi2num}) for individual
stellar mass bins. The stellar mass function is constructed solely from
the data, i.e., we do not assume any functional form for the stellar
mass function. Note that this stellar mass function is truncated at
low-mass end due to the stellar mass filter
(equation~\ref{eq:filter_sm}) at low redshifts, but at higher
redshifts the stellar mass function is truncated at higher stellar
masses due to the magnitude limit of the input galaxy catalogue. We
thus derive the lower stellar mass cutoff $M_{*,{\rm cut}}(z)$ of the
stellar mass function as a function of redshift. Then the richness
correction factor $f_N(z)$ is computed as   
\begin{equation}
f_N(z)=\frac{\int_{M_{*,{\rm cut}}(z)}^\infty d\phi/dM_{*,{\rm
      in}}(z_{\rm ref})dM_{*,{\rm in}}}
{\int_0^\infty d\phi/dM_{*,{\rm
      in}}( z_{\rm ref})dM_{*,{\rm in}}},
\label{eq:fncor}
\end{equation}
where $z_{\rm ref}$ is a reference redshift where the stellar mass
function should be sampled down to $M_l$. In this paper we adopt 
$z_{\rm ref}=0.1$. The richness correction is applied simply by
dividing the original richness by the correction factor
\begin{equation}
\hat{N}_{\rm cor}=\frac{\hat{N}_{\rm mem}}{f_N(z_{\rm cl})}.
\end{equation}
Figure~\ref{fig:fncor} shows the richness correction factor $f_N(z)$
for the SDSS cluster sample.

\subsection{Cluster catalogue}

\begin{figure}
\begin{center}
  \includegraphics[width=0.95\hsize]{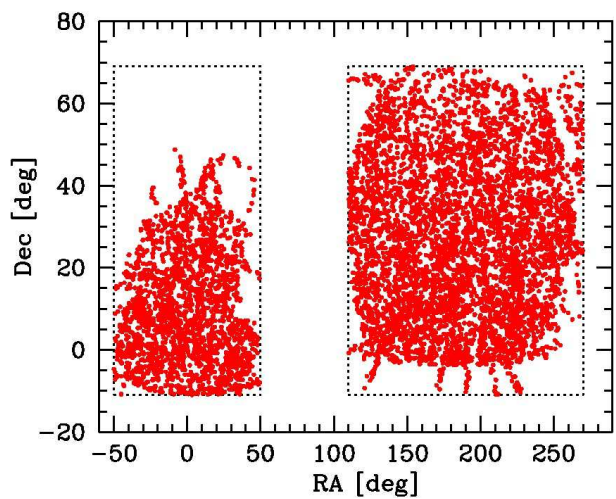}
\end{center}
\caption{Distribution of clusters with $\hat{N}_{\rm cor}>20$ and
  $0.1<z_{\rm cl}<0.3$ on the sky. Dotted lines indicate the RA and
  Dec ranges inside which our cluster catalogue is constructed.
  \label{fig:check_radec}}
\end{figure}

\begin{figure}
\begin{center}
 \includegraphics[width=0.95\hsize]{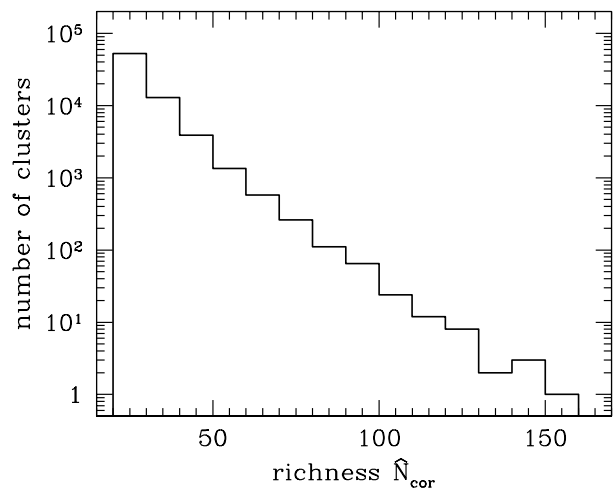}
\end{center}
\caption{The histogram of the corrected richness $\hat{N}_{\rm cor}$
  for our SDSS cluster catalogue.
  \label{fig:hist_ncor}}
\end{figure}

\begin{figure}
\begin{center}
 \includegraphics[width=0.95\hsize]{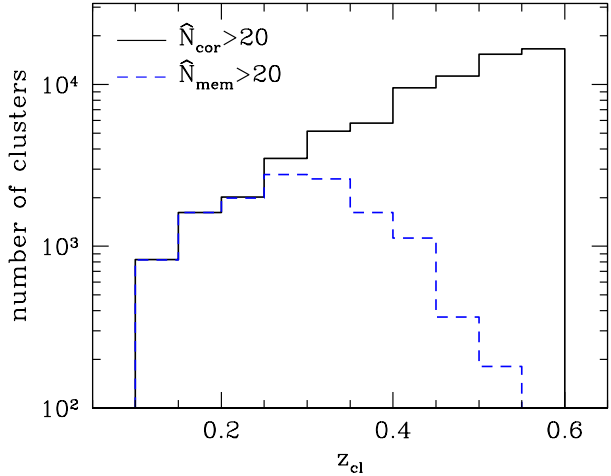}
\end{center}
\caption{The histogram of the cluster redshift $z_{\rm cl}$ for our
  SDSS cluster catalogue. The solid line shows the histogram for the
  whole cluster catalogue, whereas the dashed line is the histogram for
  clusters with the uncorrected richness $\hat{N}_{\rm mem}>20$. 
  \label{fig:zcl_dist}}
\end{figure}

We construct our SDSS DR8 cluster catalogue in the redshift range
$0.1<z_{\rm cl}<0.6$ and the richness range $\hat{N}_{\rm cor}>20$. 
The catalogue contains 71743 clusters (see
Appendix~\ref{sec:table}). The total area that satisfies the masking
criteria shown in Section~\ref{sec:masking} is $\sim 
11960$~deg$^2$. Figure~\ref{fig:check_radec}
shows the footprint of our cluster catalogue. As expected, the
spatial distribution shows large-scale structure. We also show the
richness and redshift distributions in Figures~\ref{fig:hist_ncor} 
and \ref{fig:zcl_dist}, respectively. The cluster abundance is a steep
function of the richness $\hat{N}_{\rm cor}$, which is expected from
the steep mass dependence of the cluster abundance. Before the
richness correction $f_N(z)$ is applied, the redshift distribution
begins to decrease quickly at $z\sim 0.4$ where the incompleteness of
richness estimates due to the magnitude limit of SDSS gets significant
(see Figure~\ref{fig:fncor}). After the richness correction, the
cluster number count monotonically increases out to $z\sim 0.6$, which
is qualitatively consistent with the trend expected for a
volume-limited cluster catalogue.

\begin{figure}
\begin{center}
 \includegraphics[width=0.95\hsize]{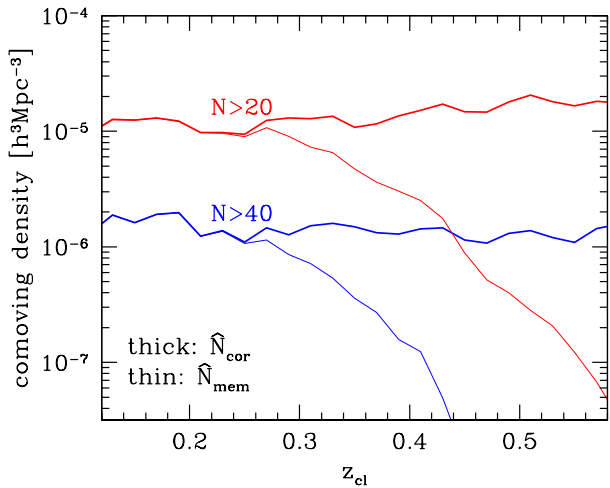}
\end{center}
\caption{The comoving number density distribution of the clusters as a
  function of cluster redshift $z_{\rm cl}$. We show the number
  densities for richness thresholds of both $20$ ({\it upper curves})
  and $40$ ({\it lower curves}). Thick lines show the number densities
  for corrected richness thresholds $\hat{N}_{\rm cor}$, and thin
  lines show those for uncorrected richness thresholds 
  $\hat{N}_{\rm mem}$. 
  \label{fig:check_zdist}}
\end{figure}

For more quantitative discussions, we compute the comoving number
density of the clusters as a function of cluster, which is shown in
Figure~\ref{fig:check_zdist}. When the uncorrected richness
$\hat{N}_{\rm mem}$ is used as a threshold, the number density indeed
rapidly decreases from $z\sim 0.4$. For the case of the corrected
richness $\hat{N}_{\rm cor}$, the number density is roughly constant
out to high redshifts, which suggests that our richness correction
method works well. We note that at low redshifts $z\la 0.35$ the
number density is quite similar to the number density of redMaPPer
clusters \citep[see][]{rykoff14}. It is worth noting that the number
density becomes slightly higher at $z\ga 0.35$, which may be explained
by increased scatters in the mass-richness relation at these high
redshifts. Because of large richness corrections, Poisson scatters in
original richness estimates at high redshifts are boosted. 

\section{Testing the performance}
\label{sec:test}

\subsection{External catalogues}
\label{sec:excatalog}

We compare the CAMIRA SDSS DR8 cluster catalogue with other cluster
catalogues in order to better understand and characterize our cluster
catalogue. For this purpose we largely follow \citet{rozo14} to adopt
several public X-ray cluster catalogues, which are briefly summarized
below. 

The XMM Cluster Survey \citep[XCS;][]{mehrtens12} is a serendipitous
search for galaxy clusters using the XMM-Newton Science Archive. The
catalogue contains 503 clusters. Spectroscopic redshifts and X-ray
temperature $T_X$ are available for nearly half of these clusters.  
We use only clusters with spectroscopic redshifts for comparisons.

The Meta-Catalogue of X-ray detected Clusters of galaxies 
\citep[MCXC;][]{piffaretti11} is based on publicly available ROSAT All-Sky
Survey \citep[RASS;][]{voges99} as well as serendipitous cluster
catalogues. There are 1559 clusters in total. In this catalogue the
X-ray luminosity $L_X$ is consistently defined in the $0.1-2.4$~keV
band integrated within $R_{500c}$, the radius within which the
interior average density becomes 500 times the critical density of the
Universe. 

The ACCEPT cluster catalogue \citep{cavagnolo09} consists of X-ray
clusters observed with {\it Chandra}. We adopt X-ray temperature $T_X$
and redshift measurements for 239 X-ray clusters from the catalogue. 

In addition, we use the spectroscopic redshifts of optical clusters
from the Sloan Giant Arcs Survey (SGAS) just for the redshift
comparison.  Specifically we use spectroscopic redshifts of 24 SGAS
clusters reported in \citet{bayliss11}, \citet{oguri12}, and
\citet{bayliss14}. For each cluster, the cluster redshift is
accurately estimated from spectroscopy of $\ga 30$ member galaxies. 

In order to compare these external catalogues with our CAMIRA cluster
catalogue, we need to match clusters between these catalogues. We
consider a simple matching criterion that clusters that are within
1~$h^{-1}{\rm Mpc}$ in the physical transverse distance and redshift
difference $\Delta z<0.1$ are matched. When there are several matching
candidates, we match clusters with smallest angular separations.  Note
that this simple matching procedure can fail in some rare cases, which
we ignore in the following analysis. 

\subsection{Cluster redshifts}

\begin{figure}
\begin{center}
 \includegraphics[width=0.95\hsize]{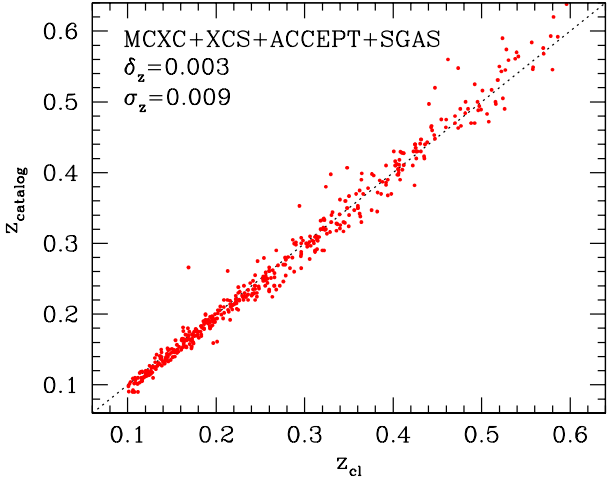}
\end{center}
\caption{Comparison between photometric cluster redshifts $z_{\rm cl}$
  (see Section~\ref{sec:refine}) and spectroscopic redshifts of
  clusters from various external catalogues. See
  Section~\ref{sec:excatalog} for descriptions of individual external
  catalogues. 
\label{fig:check_zc}}
\end{figure}

First we check the accuracy of cluster redshift $z_{\rm cl}$, which is
estimated based on photometric data only in our algorithm (see
Section~\ref{sec:refine}), with spectroscopic redshifts of clusters
from all the external catalogues described above. There are 483
clusters in total for the comparison. The result shown in
Figure~\ref{fig:check_zc} clearly indicates that our cluster redshift
is quite accurate with small outlier rate.  We quantify the accuracy
again by calculating the bias $\delta_z$ and 
scatter $\sigma_z$ of residuals $(z_{\rm cl}-z_{\rm
  catalog})/(1+z_{\rm catalog})$ with 3$\sigma$ clipping, finding 
$\delta_z=0.003$ and $\sigma_z=0.009$, which is comparable to the
accuracy achieved by, e.g., redMaPPer \citep{rozo14}.

\subsection{Comparison with X-ray properties}

The comparison between richness and X-ray properties such as X-ray
luminosity ($L_X$) and temperature ($T_X$) is useful because these X-ray
properties are thought to correlate better with cluster masses than
optical richness. This suggests that the tightness of the
mass-richness relation can be inferred from the scaling relation
between richness and X-ray properties.  For instance \citet{rykoff12}
has used this approach to refine their richness estimates. Here we
compare our richness estimates with eternal X-ray cluster catalogues
described above in a manner similar to \citet{rozo14}.

\begin{figure}
\begin{center}
 \includegraphics[width=0.95\hsize]{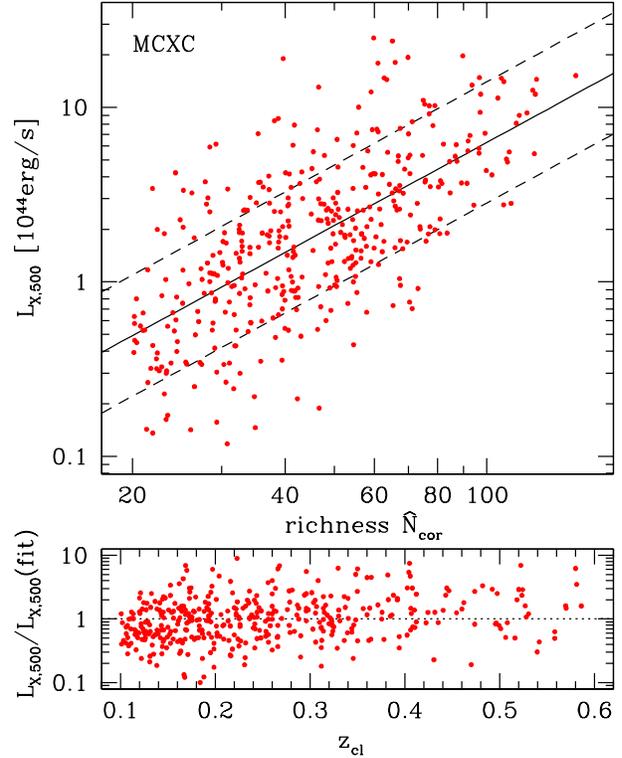}
\end{center}
\caption{Comparison between corrected richness $\hat{N}_{\rm cor}$ and
  X-ray luminosities $L_X$ for MCXC clusters. The solid and dashed lines
  show the best-fit $L_X$-$\hat{N}_{\rm cor}$ relation and 1$\sigma$
  scatter, respectively. The lower panel shows the residual of fitting
  as a function of cluster redshift $z_{\rm cl}$.
\label{fig:check_lx}}
\end{figure}

\begin{figure}
\begin{center}
 \includegraphics[width=0.95\hsize]{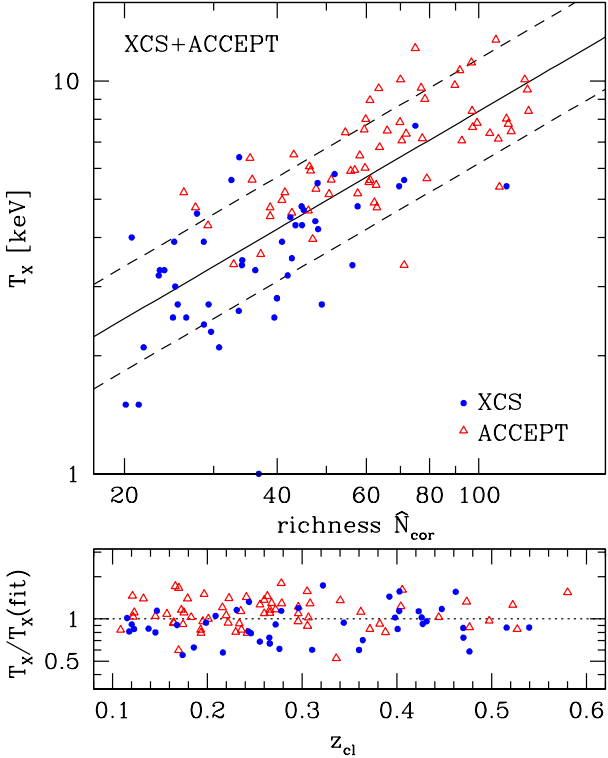}
\end{center}
\caption{Similar to Figure~\ref{fig:check_lx}, but the comparison
  between corrected richness $\hat{N}_{\rm cor}$ and X-ray
  temperatures $T_X$ is presented for XCS and ACCEPT cluster
  catalogues. In the plot clusters from XCS ({\it filled circles}) and
  ACCEPT ({\it open triangles}) are shown with different
  symbols. Lines are our fitting result for the combination of both
  the XCS and ACCEPT samples. 
\label{fig:check_tx}}
\end{figure}

Figure~\ref{fig:check_lx} compares corrected richness $\hat{N}_{\rm cor}$ and
X-ray luminosities $L_X$ for MCXC clusters. The plot shows clear positive
correlation between $\hat{N}_{\rm cor}$ and $L_X$. We fit the
relation assuming a linear relation in logarithmic space,
\begin{equation}
\log\left(\frac{L_X}{10^{44}{\rm erg\,s^{-1}}}\right)=
a_L\log\left(\frac{\hat{N}_{\rm cor}}{50}\right)+b_L,
\label{eq:N-LX}
\end{equation}
using the least square method. We find the best-fit slope $a_L=1.58\pm
0.09$ and normalization $b_L=0.32\pm 0.02$. The 1$\sigma$ scatter in
$\log L_X$ is 0.35 without any outlier rejections. 
Figure~\ref{fig:check_lx} indicates that  residuals of the fitting
show no strong correlation with cluster redshift. 

Figure~\ref{fig:check_tx} shows a similar comparison for X-ray
temperature $T_X$ for XCS and ACCEPT clusters. Again, $T_X$ correlates
well with richness $\hat{N}_{\rm cor}$. Assuming the scaling relation
of the form 
\begin{equation}
\log\left(\frac{T_X}{\rm keV}\right)=
a_T\log\left(\frac{\hat{N}_{\rm cor}}{50}\right)+b_T,
\label{eq:N-TX}
\end{equation}
we find the best-fit slope $a_T=0.76\pm0.06$ and normalization
$b_T=0.70\pm0.01$ with 1$\sigma$ scatter 0.13. Again residuals show no
strong trend with cluster redshift.

As discussed in \citet{rozo14}, there is a systematic offset between
X-ray temperatures of XCS and ACCEPT clusters. This is also evident
from Figure~\ref{fig:check_tx}, which indicates that ACCEPT
clusters appear to have larger X-ray temperatures for a given
richness. We fit each cluster sample to equation~(\ref{eq:N-TX}), and
find $a_T=0.61\pm 0.13$, $b_T=0.62\pm0.03$, and the scatter of 0.14
for XCS clusters, and $a_T=0.51\pm0.07$ and $b_T=0.76\pm0.01$, and the
scatter of 0.10 for ACCEPT clusters. Our result indicates $\approx
40$\% systematic offset of X-ray temperatures, which is consistent
with \citet{rozo14}.  \citet{rozo14} argued that the temperature
offset can be ascribed to differences of X-ray temperature
definitions between ACCEPT and XCS clusters, and therefore is not
problematic. 

In comparison with results presented in \citet{rozo14}, we find that
the CAMIRA cluster catalogue and richness estimate are comparable to
the redMaPPer cluster catalogue in terms of the tightness of cluster
richness with X-ray properties. The scatter in the scaling relations
translates into the scatter of the mass-richness relation of
$\sigma_{\ln M}\sim 0.3-0.4$ (i.e., scatter of $\sim 0.15$ for $\log
M$). While the slight increase of the comoving number density at
higher redshift (see Figure~\ref{fig:check_zdist}) implies enhanced
scatter of the mass-richness relation at $z\ga 0.35$, it is not very
clear in this analysis using X-ray. In fact the scatter may be
affected by the incompleteness of X-ray catalogues we use for the
comparisons.  X-ray data are available only for massive clusters,
which is particularly true for high-redshift clusters, and hence less
X-ray luminous clusters are not included in deriving the scaling
relation. This Malmquist bias can lead to an underestimation of
scatters as well as systematic shifts of mean relations. Therefore
more careful comparisons with X-ray properties should take account of
the sample incompleteness, which we leave for future work.

For a further test, we also conduct the ``X-ray mass scatter''
analysis presented in \citet{rozo14}. This is done by scrambling the
richness values for the matched cluster catalogue, re-fit the richness
temperature relation, and derive the scatter for this shuffled
catalogues. We create 1000 realizations of the shuffled cluster
catalogues for both XCS and ACCEPT clusters. We find average scatters
of 0.17 and 0.13 for XCS and ACCEPT clusters, respectively, which
should be compared with 0.14 and 0.10 for unshuffled XCS and ACCEPT
clusters, respectively. Therefore the scatter is indeed reduced
relative to the shuffled richness catalogue. For all the shuffled
cluster samples, their scatters are larger than those of the
unshuffled cluster samples, which indicate that the reduction of the
scatter is more than $3\sigma$ significant. This is again comparable
to the performance of redMaPPer \citep[see][]{rozo14}. 

\begin{figure}
\begin{center}
 \includegraphics[width=0.95\hsize]{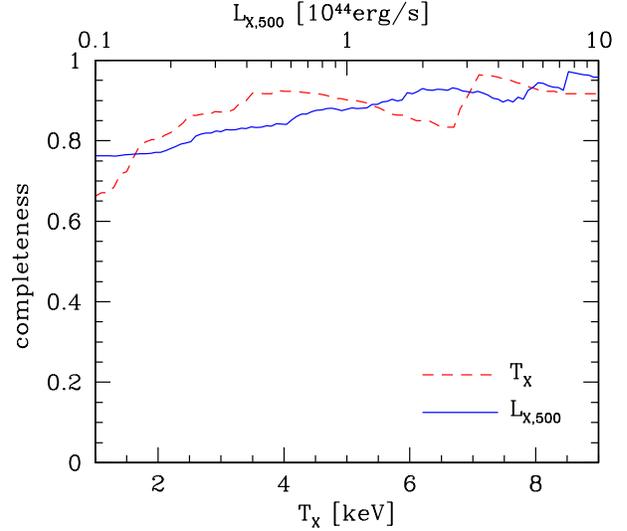}
\end{center}
\caption{Completeness as a function of X-ray temperature 
({\it dashed}) or luminosity ({\it solid}) threshold, for matched
  cluster sample at $0.1<z<0.5$. In order to compare the result with
  that presented in \citet{rozo14}, here we use an approximate
  treatment to determine which X-ray clusters fall within the optical
  mask, which leads to an underestimate of the completeness.  
\label{fig:compl}}
\end{figure}

\subsection{Completeness}

Here we conduct a simple completeness estimate using the external
X-ray catalogues, following the procedure given in \citet{rozo14}.
First we need to know whether any given X-ray clusters fall within the
footprint of the CAMIRA SDSS DR8 catalogue. We adopt an approximated
approach that X-ray clusters that fall within $40'$ of any clusters in
the CAMIRA SDSS DR8 catalogue are within the footprint. Note that this
procedure tends to underestimate the completeness, but this allows us
to compare our result with that presented in \citet{rozo14}. Then we
derive the completeness as a function of X-ray luminosity and
temperature thresholds. We use clusters in the redshift range
$0.1<z<0.5$, the same range adopted in \citet{rozo14}.   

Figure~\ref{fig:compl} show the completeness from cross-matching with
XCS, ACCEPT, and MCXC clusters. We find that the completeness is quite
high, $>0.9$ for X-ray luminous and high temperature clusters. The
high completeness is indeed comparable to redMaPPer result
\citep[see][]{rozo14}. The completeness is less than unity at the high
$T_X$ and $L_X$ end, due to the approximate treatment to determine
which X-ray clusters fall within the optical mask as mentioned above
\citep[see also Figure 9 of][]{rozo14}.

In this analysis we did not explicitly match the comoving volume
density. At low redshifts, $z\la 0.35$, the comoving number densities
of CAMIRA and redMaPPer cluster catalogues are similar, but at higher
redshifts the CAMIRA cluster catalogue has much higher number density
of clusters than redMaPPer. However, we note that even if we restrict
the redshift range to $0.1<z<0.3$ the completeness is similar to that
plotted in Figure~\ref{fig:compl}. 

\begin{figure}
\begin{center}
 \includegraphics[width=0.95\hsize]{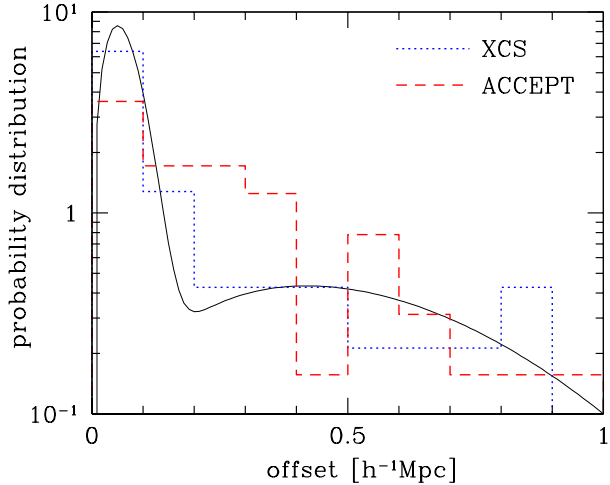}
\end{center}
\caption{Distributions of the offset between CAMIRA and X-ray
  cluster centres. The offset distributions are computed both for
  XCS ({\it dotted}) and ACCEPT ({\it dashed}) clusters. The solid
  line shows a simple two-component model with the fraction of centred
  component $f_{\rm cen}=0.7$, the standard deviation of the centred
  component $R_{s,{\rm cen}}=0.05h^{-1}{\rm Mpc}$, and the standard
  deviation of the miscentred component $R_{s}=0.42h^{-1}{\rm Mpc}$
  (see equation~\ref{eq:offset}). 
\label{fig:offdist}}
\end{figure}

\subsection{Offsets between optical and X-ray cluster centres}

Finding cluster centres is one of the most important challenges in
optical cluster finding algorithms. As X-ray emissions trace the
gravitational potential of galaxy clusters, comparisons of centres of
optically selected clusters with X-ray centres of same clusters
provide a useful means of testing the accuracy of the centring
algorithm of optical cluster finders. 

As in \citet{rozo14}, for each matched X-ray cluster we compute a
physical transverse distance between optical and X-ray cluster centres to
derive the offset distribution. Figure~\ref{fig:offdist} shows the
offset distributions both for XCS and ACCEPT clusters. We find that the
offset is generally small, but the distribution has a long tail to
large offsets. The distribution of the offset $R$ is often modelled by
the two component Gaussian distributions
\citep[e.g.,][]{johnston07,oguri11a} 
\begin{eqnarray}
p(R)&=&f_{\rm cen}\frac{R}{R_{s,{\rm
      cen}}^2}\exp\left(-\frac{R^2}{2R_{s,{\rm cen}}^2}\right)\nonumber\\
&&+(1-f_{\rm cen})\frac{R}{R_{s}^2}\exp\left(-\frac{R^2}{2R_{s}^2}\right).
\label{eq:offset}
\end{eqnarray}
Figure~\ref{fig:offdist} plot an example of the two-component model
with $f_{\rm cen}=0.7$, $R_{s,{\rm cen}}=0.05h^{-1}{\rm Mpc}$, and
$R_{s}=0.42h^{-1}{\rm Mpc}$, which very roughly explains to the
observed distribution. Given the critical importance of miscentring
of optical cluster samples for, e.g., stacked weak lensing analysis
(see below), it is important to examine the offset more carefully
using larger samples of X-ray clusters.

\section{Weak lensing mass calibration}
\label{sec:weaklens}

\subsection{Measurement}

Here we employ a public shear catalogue of the CFHTLenS
\citep{heymans12} to calibrate the mass-richness relation. The
CFHTLenS shear catalogue is based on a sophisticated Bayesian galaxy
shape measurement with careful calibrations using simulated galaxy
images \citep{miller13}. The photometric redshift estimate is also
available for each galaxy \citep{hildebrandt12}. Readers are referred
to \citet{erben13} for more details of the public shear catalogue. 

The shear catalogue provides two ellipticity components ($e_1$, $e_2$)
in the celestial coordinate system as well as multicative and additive 
calibration factors $m$ and $c_2$ for each galaxy. For each reference
centre, we compute the tangential ellipticity component as
\begin{equation}
e_+=-e_1\cos 2\phi-(e_2-c_2)\sin 2\phi,
\end{equation}
where $\phi$ is the angle of the position of the source galaxy in the
polar coordinate system, measured counterclockwise from West. Then we
measure the average projected mass density in each radial bin $R$
(physical units) as 
\begin{equation}
\Delta\Sigma(R)=\frac{\sum_i w_i\Sigma_{{\rm cr},i}e_{+,i}}{\sum_i (1+m_i)w_i},
\end{equation}
where $\Sigma_{\rm cr}$ is the critical surface mass density for
lensing, computed from the cluster redshift $z_{\rm cl}$ and the
photometric redshift $z_{p,{\rm best}}$ of the source galaxy. The
index $i$ runs over source galaxies in the specified radial bin behind
all foreground lensing clusters considered. We choose the weight $w_i$
as  
\begin{equation}
w_i=\frac{w_{g,i}}{\Sigma_{{\rm cr},i}^2},
\end{equation}
where $w_g$ is the weight factor of each source galaxy provided by the
CFHTLenS shear catalogue. The critical surface density is introduced
in the weight factor to downweight source galaxies whose redshifts are
close to lens redshifts and therefore their weak lensing effects are
inefficient \citep[see, e.g.,][]{mandelbaum13}. 

One of the most important potential systematic effects in cluster weak
lensing analysis is the dilution effect by cluster member galaxies 
\citep[e.g.,][]{medezinski07,okabe13}. One can mitigate the dilution
effect by selecting appropriate background galaxies for weak lensing
analysis. While photometric redshifts are available for individual source
galaxies in the CFHTLenS catalogue, imperfect photometric redshifts
lead to the contamination of cluster member galaxies in the source
galaxy sample, even if the photometric redshift cut is applied to
select only background galaxies. In this paper, we adopt the following
procedure to construct a secure background galaxy sample. For each
galaxy, the CFHTLenS shear catalogue provides the best photometric
redshift estimate $z_{p,{\rm best}}$ as well as the full probability
distribution function (PDF) of the photometric redshift, $P(z_p)$. We
then define a subsample of source background galaxies by
\begin{equation}
\int_{z_{p,{\rm min}}}^\infty P(z_p)dz_p>p_{\rm cut},
\label{eq:zpmin}
\end{equation}
with $p_{\rm cut}=0.98$, and  
\begin{equation}
z_{p,{\rm best}}<1.3.
\end{equation}
We adopt $z_{p,{\rm min}}$ to be 0.05 higher than the upper limit of
the cluster redshift bin of interest. The first condition assures that 
the PDF does not extend down to cluster redshifts, and therefore
should be able to select background galaxies more securely than simple
photometric redshift cuts based only on $z_{p,{\rm best}}$. The second
cut is included because photometric redshift estimates of galaxies
with $z_{p,{\rm best}}>1.3$ are thought to be less secure 
\citep[see, e.g.,][]{kilblinger13}. 

To test the validity of our approach to select the background galaxy
sample, we check the average number density of background galaxies as
a function of distance from cluster centres. We find that the average
number density is nearly flat for low richness cluster samples,
whereas the number density decreases toward the centre for high
richness clusters, which can be explained by the lensing magnification
as well as obscuration by cluster member galaxies. The lack of
increase of the background galaxy number density toward the cluster
centre assures that our background galaxy sample is not significantly
contaminated by cluster member galaxies.  

\subsection{Model}
\label{sec:nfwmodel}

We fit the stacked weak lensing signal with our theoretical model to
extract cluster parameters. The mass distribution of individual
clusters are assumed to follow the \citet[][hereafter NFW]{navarro97}
density profile. We assume that the BCG selected in our algorithm to
each cluster centre, but some BCGs defined in our algorithm may in
fact correspond to satellite galaxies rather than galaxies in halo
centres. We include this miscentring effect in the
stacked weak lensing profile using the Fourier space approach
developed by \citet{oguri11a}. For the miscentring model, we adopt a
two-component model that was also used by \citet{oguri11a}. This
model, with the explicit form presented in equation~(\ref{eq:offset}),
assumes that one component is well centred and the other component 
whose offset PDF is described by the two-dimensional Gaussian
distribution.  \citet{oguri11a} found that the average convergence
profile with the miscentring effect in the angular Fourier space is
described by 
\begin{equation}
\kappa_{\rm NFW, off}(\ell)=\kappa_{\rm NFW}(\ell)\left[f_{\rm
    cen}+(1-f_{\rm cen})\exp\left(-\frac{1}{2}\sigma_s^2\ell^2\right)\right],
\end{equation}
where $f_{\rm cen}$ is the fraction of the well-centred cluster
component and $\sigma_s=R_s/D_A(z)$ specifies the size of the offset
PDF. Here we assumed $R_{s,{\rm cen}}\approx 0$ in
equation~(\ref{eq:offset}) for simplicity. For the
original NFW profile in the Fourier space \citep[see][]{oguri11a},  
in fact we employ the Fourier transform of a truncated NFW profile
\citep{baltz09} presented in \citet{oguri11b}, but choose the
truncation radius sufficiently large to describe the untruncated NFW
profile. 

The concentration parameter $c_{\rm vir}=r_{\rm vir}/r_s$ is an
important parameter that quantifies the mass concentration of the NFW
profile. We assume the following mass and redshift dependences
\citep{duffy08} 
\begin{equation}
c_{\rm vir}=c_{\rm norm}\frac{7.85}{(1+z)^{0.71}}\left(
\frac{M_{\rm vir}}{2\times 10^{12}h^{-1}M_\odot}\right)^{-0.081},
\end{equation}
and treat the overall normalization $c_{\rm norm}$ as a parameter in
order to take account of the uncertainty of the concentration
parameter. 

Given the Fourier space description of the NFW profile, the convergence
and tangential shear profiles are computed as 
\begin{equation}
\kappa_{\rm off}(R)=\int \frac{\ell d\ell}{2\pi}\kappa_{\rm
  NFW,off}(\ell)J_0(\ell R/D_A(z)),
\end{equation}
\begin{equation}
\gamma_{+,{\rm off}}(R)=\int \frac{\ell d\ell}{2\pi}\kappa_{\rm
  NFW,off}(\ell)J_2(\ell R/D_A(z)),
\end{equation}
where $J_0(x)$ and $J_2(x)$ are zero-th and second order Bessel
functions. Since weak lensing in fact measures the reduced shear, 
we approximately compute the surface mass density for a fixed halo mass
as 
\begin{equation}
\Delta\Sigma(R)=\frac{\Sigma_{\rm cr}\gamma_{+,{\rm
      off}}(R)}{1-\kappa_{\rm off}(R)},
\label{eq:redsh}
\end{equation}
where $\Sigma_{\rm cr}$ is computed using the mean lens and source
redshifts of the sample. 

\subsection{Fitting procedure}

We consider two redshift slices for the stacked weak lensing analysis,
the low-redshift slice with $0.1<z_{\rm cl}<0.3$ and the high-redshift 
slice with $0.4<z_{\rm cl}<0.6$. We consider these two redshift bins
given the possible change of cluster properties in our cluster sample
at $z\sim 0.35$ (see, e.g., Figure~\ref{fig:check_zdist}).
We set the photometric redshift cut
(see equation~\ref{eq:zpmin}) $z_{p,{\rm min}}=0.35$ and $0.65$ for
the low- and high-redshift cluster samples, respectively.  For each
redshift slice we consider four richness bins defined by
$20<\hat{N}_{\rm cor}<25$, $25<\hat{N}_{\rm cor}<35$, 
$35<\hat{N}_{\rm cor}<50$, and $50<\hat{N}_{\rm cor}<90$. Thus there
are 8 subsamples in total for the stacked weak lensing analysis. We
only use clusters in the overlapping regions ($\sim 120$~deg$^2$) of
SDSS DR8 and CFHTLenS.  

The radial range of profile fitting must be chosen carefully to reduce
various systematic errors. For instance, shear signals near the halo
centre are difficult to interpret for several reasons. First, our
calculation of reduced shear given in equation~(\ref{eq:redsh})
involves an approximation which becomes less accurate toward the halo
centre. Second, the dilution effect of cluster member galaxies is more
pronounced near the centre, so any residual contamination of cluster
member galaxies in the source shear catalogue, if exists, decreases
tangential shear signals. Third, the CFHTLenS shear measurement is
less tested again simulated galaxy images in the high shear regime
like cluster centres. As shown in \citet{becker11} and
\citet{oguri11b}, the maximum radius of fitting is also important for
unbiased measurement, because of increasing contributions of the
so-called two-halo term to the stacked weak lensing profile. Also
stacking around random points suggests that there appears to be
residual systematics in the CFHTLenS shear measurement at large radii,
$R\ga 10h^{-1}{\rm Mpc}$ \citep{miyatake14,covone14}. Thus we
conservatively choose the radial range of our profile fitting to
$0.158<R/(h^{-1}{\rm Mpc})<2.09$, which is divided into 7
logarithmically spaced bins with an interval of $\Delta(\log R)=0.16$.
For simplicity we do not consider the cosmic shear error, as the
cosmic shear error is subdominant in the radius range considered
here \citep[see][]{miyatake14}. 

Parameters of our shear profile model includes the halo mass
$\langle M_{\rm vir}\rangle$, the normalization of the concentration
parameter $c_{\rm norm}$, the fraction of the centred component
$f_{\rm cen}$, the miscentring size $R_s$. Since the miscentring
parameters degenerate with the concentration parameter
\citep{oguri11a}, we add a conservative Gaussian prior to 
$\log c_{\rm norm}$ as $\log c_{\rm norm}= 0\pm 0.2$, and fix the
miscentring size to $0.42h^{-1}{\rm Mpc}$ based on the analysis
result of the mock galaxy catalogue in \citet{johnston07}. 
Therefore the number of degree of freedom of our fitting is 5. 

\begin{figure}
\begin{center}
 \includegraphics[width=0.95\hsize]{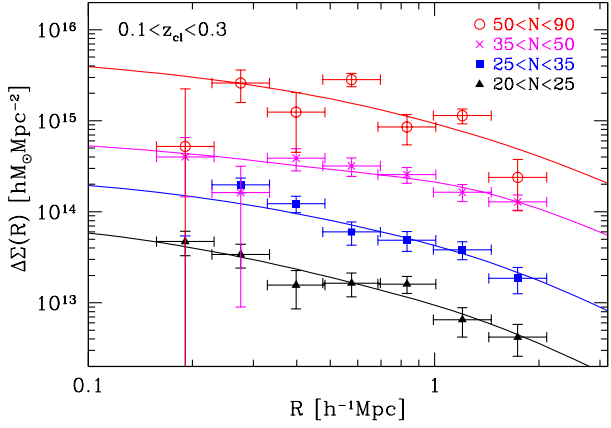}
\end{center}
\caption{Stacked surface mass density profiles from the CAMIRA SDSS
  DR8 clusters and CFHTLenS shear catalogue. Here we plot the low
  cluster redshift bin defined by $0.1<z_{\rm cl}<0.3$. Different
  symbols show results for different richness bins; 
  $20<\hat{N}_{\rm cor}<25$ ({\it filled triangles}), 
  $25<\hat{N}_{\rm cor}<35$ ({\it filled squares}), 
  $35<\hat{N}_{\rm cor}<50$ ({\it crosses}), and 
  $50<\hat{N}_{\rm cor}<90$ ({\it open circles}), 
  which are shifted vertically by $-0.5$, 0, $0.5$, and 1~dex,
  respectively, for illustrative purpose. Solid lines show
  best-fitting NFW profiles including the miscentring effect (see
  Section~\ref{sec:nfwmodel} for more details). 
\label{fig:ds_stack_lowz}}
\end{figure}

\begin{figure}
\begin{center}
 \includegraphics[width=0.95\hsize]{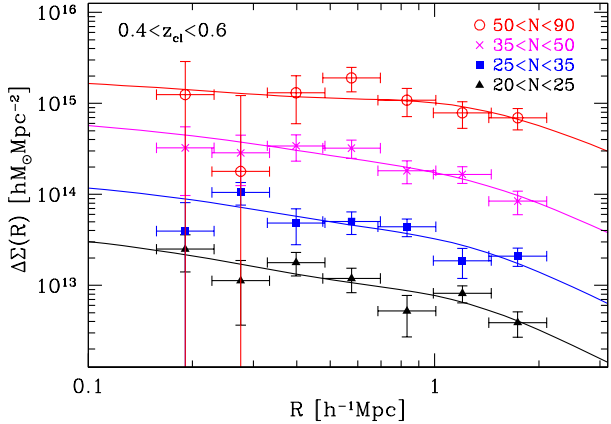}
\end{center}
\caption{Similar to Figure~\ref{fig:ds_stack_lowz}, but results for
  the high cluster redshift bin $0.4<z_{\rm cl}<0.6$ are shown.
\label{fig:ds_stack_highz}}
\end{figure}

\subsection{Results}

Stacked surface mass density profiles for low ($0.1<z_{\rm cl}<0.3$)
and high ($0.4<z_{\rm cl}<0.6$) redshift cluster samples are shown in
Figures~\ref {fig:ds_stack_lowz} and \ref{fig:ds_stack_highz},
respectively. It is clear that stacked weak lensing signals are
detected significantly for all the 8 cluster subsamples. As expected,
the signals decreases with increasing projected radius, which are
found to be fitted reasonably well by our model including the
miscentring effect (equation~\ref{eq:redsh}). From the comparisons
with the theoretical model we derive constraints on model parameters
such as the mean halo mass $\langle M_{\rm vir}\rangle$ and 
the fraction of the centred component $f_{\rm cen}$. We summarize the
results in Table~\ref{tab:fit}.

\begin{figure}
\begin{center}
 \includegraphics[width=0.95\hsize]{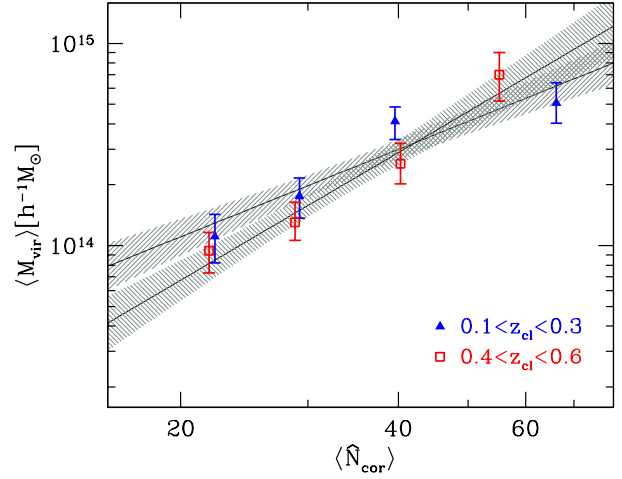}
\end{center}
\caption{Scaling relations between the mean richness 
 $\langle\hat{N}_{\rm cor}\rangle$ and the mean halo mass 
 $\langle M_{\rm vir}\rangle$ inferred from the CFHTLenS stacked weak
  lensing analysis. Filled triangles show the relation for
  the low-redshift ($0.1<z_{\rm cl}<0.3$) cluster sample, whereas open
  squares show the relation for the high-redshift 
  ($0.4<z_{\rm cl}<0.6$) cluster sample (see also
  Table~\ref{tab:fit}). Solid lines with shading are power-law fits
  (equation~\ref{eq:plfit_mn}) and $1\sigma$ error of the scaling
  relations.   
\label{fig:massrich}}
\end{figure}

The fitting results clearly indicate that the mean halo mass inferred
from stacked weak lensing correlates well with the richness. To
illustrate this point, we show the scaling relation in
Figure~\ref{fig:massrich}. We do not find significant difference in
the scaling relations between the low- and high-redshift clusters.
Our result indicates that the richness limit of $\hat{N}_{\rm cor}>20$
for the CAMIRA SDSS DR8 catalogue corresponds to the cluster virial mass
limit of $M_{\rm vir}\ga 1\times 10^{14}h^{-1}M_\odot$ over the
redshift range of $0.1<z_{\rm cl}<0.6$. The virial mass limit may be
slightly lower at higher redshifts, possibly due to the increased
scatter of the richness estimate as discussed above. 

\begin{table*}
 \caption{Results of stacked weak lensing analysis with the CFHTLenS
   shear catalogue. 
\label{tab:fit}}   
 \begin{tabular}{@{}ccccccccc}
 \hline
     $\hat{N}_{\rm cor}$ range
   & $z_{\rm cl}$ range
   & $N_{\rm cluster}$
   & $\langle\hat{N}_{\rm cor}\rangle$
   & $\langle z_{\rm cl}\rangle$
   & $\langle z_{p,{\rm best}}\rangle$
   & $\log (\langle M_{\rm vir}\rangle/h^{-1}M_\odot)$
   & $\log c_{\rm norm}$$^a$
   & $f_{\rm cen}$\\
  \hline
 20--25 & 0.1--0.3 &  35 & 22.3 & 0.25 & 0.86 & $14.04_{-0.13}^{+0.11}$ & $ 0.01_{-0.19}^{+0.20}$ & $0.77_{-0.25}^{+0.23}$ \\ 
 25--35 & 0.1--0.3 &  29 & 29.2 & 0.26 & 0.86 & $14.24_{-0.11}^{+0.09}$ & $-0.09_{-0.16}^{+0.17}$ & $0.83_{-0.26}^{+0.17}$ \\ 
 35--50 & 0.1--0.3 &  14 & 39.6 & 0.23 & 0.86 & $14.62_{-0.09}^{+0.07}$ & $-0.04_{-0.18}^{+0.20}$ & $0.49_{-0.19}^{+0.24}$ \\ 
 50--90 & 0.1--0.3 &   4 & 66.2 & 0.17 & 0.86 & $14.71_{-0.10}^{+0.10}$ & $-0.03_{-0.11}^{+0.15}$ & $0.99_{-0.29}^{+0.01}$ \\ 
 20--25 & 0.4--0.6 & 291 & 21.9 & 0.51 & 0.98 & $13.97_{-0.11}^{+0.09}$ & $ 0.01_{-0.21}^{+0.17}$ & $0.39_{-0.12}^{+0.21}$ \\ 
 25--35 & 0.4--0.6 & 186 & 28.8 & 0.51 & 0.98 & $14.12_{-0.09}^{+0.10}$ & $-0.04_{-0.19}^{+0.18}$ & $0.47_{-0.17}^{+0.21}$ \\ 
 35--50 & 0.4--0.6 &  72 & 40.3 & 0.50 & 0.98 & $14.40_{-0.10}^{+0.10}$ & $-0.04_{-0.18}^{+0.20}$ & $0.55_{-0.20}^{+0.27}$ \\ 
 50--90 & 0.4--0.6 &  13 & 55.2 & 0.52 & 0.98 & $14.85_{-0.13}^{+0.11}$ & $-0.03_{-0.18}^{+0.20}$ & $0.33_{-0.18}^{+0.28}$ \\ 
 \hline
 \end{tabular}
\flushleft{$^a$ Note that the Gaussian prior $\log c_{\rm
    norm}=0\pm0.2$ is included in fitting.}
\end{table*}

We quantify the mean mass-richness relations by fitting them to the
following power-law relation
\begin{equation}
\log\left(\frac{\langle M_{\rm vir}\rangle}{h^{-1}M_\odot}\right) =
a_M\log\left(\frac{\langle\hat{N}_{\rm cor}\rangle}{30}\right)+b_M.
\label{eq:plfit_mn}
\end{equation}
We find $a_M=1.44\pm0.27$ and $b_M=14.30\pm 0.05$ for the low redshift
cluster sample, and $a_M=2.10\pm0.39$ and $b_M=14.20\pm 0.06$ for the
high redshift cluster sample. The best-fit relations are shown in
Figure~\ref{fig:massrich}.

In addition to the mean mass-richness relation, the stacked weak
lensing analysis provides some insight into the halo miscentring
effect. Although our constraints on the miscentring parameter $f_{\rm
  cen}$ (see Table~\ref{tab:fit}) is not tight due to the degeneracy
with the concentration parameter, we see a trend that $f_{\rm
  cen}$ is smaller at higher redshifts. In particular $f_{\rm cen}$
is most significantly smaller than unity for high redshift,
low-richness clusters. This is presumably due to the fact that
these clusters intrinsically contain small number of cluster member
galaxies, and therefore proper selections of central galaxies may be
more challenging. Our result here is another example of how weak
lensing can be used to study halo miscentring effects
\citep{oguri10,george12,ford14}.  

\begin{figure}
\begin{center}
 \includegraphics[width=0.95\hsize]{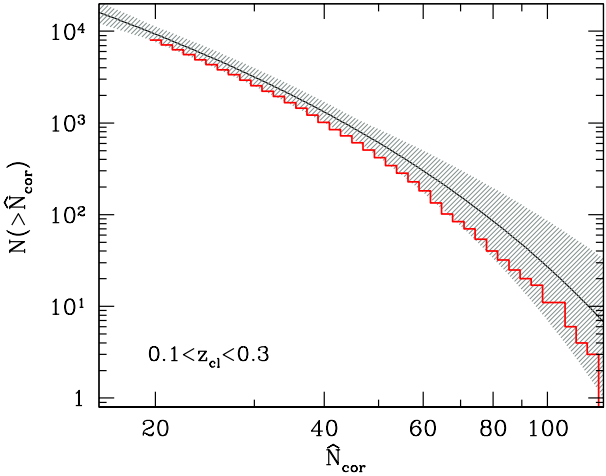}
\end{center}
\caption{The cumulative number distribution of clusters as a function
  of the corrected richness $\hat{N}_{\rm cor}$ for the entire SDSS
  DR8 footprint. Here we consider the cluster redshift range
  $0.1<z_{\rm cl}<0.3$ where our richness estimates are more secure. 
  The histogram shows the distribution in our CAMIRA cluster sample. 
  The solid line shows the theoretical expectation from the
  mass-richness scaling relation of equation~(\ref{eq:plfit_mn}). The
  shaded region represents $1\sigma$ range from the statistical
  uncertainty in the mass-richness scaling relation (see
  Figure~\ref{fig:massrich}).  
\label{fig:ncum}}
\end{figure}

Finally we perform a simple test to compare the observed cluster
abundance with the theoretical expectation. Specifically we adopt the
power-law scaling relation obtained from the CFHTLenS stacked weak
lensing analysis (equation~\ref{eq:plfit_mn}) to convert a halo mass
function \citep{sheth99} to the number density distribution of
clusters as a function of the corrected richness $\hat{N}_{\rm cor}$. 
The number density distribution is integrated over a specific cluster
redshift range to obtain the expected number distribution in the entire
SDSS DR8 footprint assuming he total area of $11960$~deg$^2$. We show
a tentative comparison of the cumulative number density distributions in
Figure~\ref{fig:ncum}. The Figure indicates that the abundance in our
CAMIRA cluster catalogue is in good agreement with the theoretical
expectation from the weak lensing calibrated mass-richness relation.  
We however note that for more careful comparisons we should take
account of the distribution as well as scatter of the mass-richness
relation \cite[see, e.g.,][]{oguri11a}. Understanding the shape of the
PDF of the mass-richness relation is a key for robust constraints on
cosmological parameters from the cluster abundance analysis, which we
will address in future work. It
is worth noting that the cosmological parameter dependence of stacked
weak lensing analysis to calibrate  the mass-richness relation must
also be taken into account for the cosmological analysis \citep{more13}. 

\section{Summary}
\label{sec:summary}

In this paper, we have presented a new cluster finding algorithm,
CAMIRA, which identifies the concentration of red-sequence galaxies in
photometric surveys. The algorithm makes use of the SPS model to
predict red-sequence galaxy colours for an arbitrary set of filters. 
The model must be calibrated using a sample of spectroscopic
red-sequence galaxies in order to achieve enough accuracy necessary
for various applications. For a given redshift we count the number of
red-sequence galaxies at that redshift, where the ``number'' of
individual galaxies is a smooth function of $\chi^2$ of fitting to the
SPS model, using a spatial filter that is designed to subtract the
background level. We also restrict the stellar mass range using a
smooth filter function. We identify cluster candidates by locating
peaks in the three-dimensional richness map, and each cluster
candidate is refined by iteratively finding the BCG and best-fit
cluster redshift. In addition the algorithm takes proper account of
masking effects. 

We have applied the algorithm to the SDSS DR8 imaging data covering
$\sim 11960$~deg$^2$. We have first calibrated the SPS model using a
large sample of spectroscopic galaxies in SDSS and BOSS. We have
constructed a catalogue containing 71743 clusters in the redshift
range of $0.1<z_{\rm cl}<0.6$ with the corrected richness
$\hat{N}_{\rm cor}>20$. The number of clusters per redshift bin
increases with increasing cluster redshift, simply because the
comoving volume increases. The comoving number density is roughly
constant over the entire cluster redshift range. 

We have compared the CAMIRA SDSS DR8 cluster catalogue with external
cluster catalogues to test its performance. The comparison of our
photometric cluster redshift estimates with spectroscopic cluster
redshifts from the external catalogues indicates that the photometric
cluster redshift is accurate with low bias and scatter. We have also
compared the corrected richness with X-ray luminosities and
temperatures and found good correlations. Scatters of these relations
are comparable to those found in other optical SDSS cluster samples
such as redMaPPer.

We have derived stacked weak lensing signals for the SDSS cluster
catalogue using the public CFHTLenS shear catalogue. Despite the small
overlapping area of $\sim 120$~deg$^2$, we have detected stacked
lensing signals significantly for all the 8 subsamples divided by
redshift and richness. The mean halo mass inferred from the lensing
analysis clearly correlates with richness. There is no significant
difference of the mass-richness relations between low ($0.1<z_{\rm
  cl}<0.3$) and high ($0.4<z_{\rm cl}<0.6$) redshift cluster samples. 
The stacked weak lensing analysis indicates that the richness limit of
$\hat{N}_{\rm cor}>20$ for the SDSS cluster catalogue corresponds to
the cluster virial mass limit of about  $M_{\rm vir}\ga 1 \times
10^{14}h^{-1}M_\odot$. We have also obtained constraints on
miscentring from the stacked weak lensing analysis. At the low
redshift our results are consistent with no miscentring component
($f_{\rm cen}=1$), while miscentring appears to be significant for
the high redshift clusters. The cluster abundance is found to be
consistent with theoretical expectation obtained using the
mass-richness relation calibrated by weak lensing, though for more
careful comparisons we need to take account of the scatter of the
mass-richness relation.   

This cluster finding algorithm is developed with the application to
ongoing and future wide-field optical imaging surveys \citep[in
  particular Subaru Hyper Suprime-Cam;][]{miyazaki12} in mind.
In these future surveys, imaging data are much deeper than SDSS,
allowing us to detect all the member galaxies of interest out to very
high redshifts, $z\sim 1$. In addition we will be able to obtain
better stacked weak lensing signals to examine the mass-richness
relation more extensively. Furthermore, careful comparisons with
cluster catalogues in other wavelength such as X-ray and SZ are
important in understanding the mass-richness relation as well as
miscentring effects. We can obtain independent information on the
mass-richness relation from clustering analysis, such as
the auto-correlation function and large-scale stacked weak lensing
profile, which helps constraining the mass-richness relation further. 
These are necessary steps for turning cluster of galaxies into a
useful probe of cosmology. 

\section*{Acknowledgments}
I thank members of the HSC cluster working group, especially Masayuki
Tanaka, Yen-Ting Lin, Masahiro Takada, and Surhud More, for fruitful
discussions, and Eduardo Rozo for useful comments.  
I also thank an anonymous referee for various suggestions.
This work was supported in part by the FIRST program 
``Subaru Measurements of Images and Redshifts (SuMIRe)'', World
Premier International Research Center Initiative (WPI Initiative),
MEXT, Japan, and Grant-in-Aid for Scientific Research from the JSPS 
(26800093). 


\appendix
\section{CAMIRA SDSS DR8 Cluster Catalogue}
\label{sec:table}

\begin{table}
 \caption{The CAMIRA SDSS DR8 cluster catalogue. The full table in the
   online edition only. 
\label{tab:catalogue}}   
 \begin{tabular}{@{}ccccc}
 \hline
     RA (J2000)
   & Dec. (J2000)
   & $z_{\rm cl}$
   & $\hat{N}_{\rm cor}$
   & $\hat{N}_{\rm mem}$\\
  \hline
  0.000095 & 24.902249  & 0.4832 &  21.515 &  8.407 \\
  0.009577 &   5.288260 & 0.1761 &  29.235 &  29.164 \\
  0.012499 &  34.580621 & 0.3100 &  22.468 &  18.512 \\
  0.013423 &  22.861665 & 0.5140 &  21.928 &   7.573 \\
  0.013767 &  31.231751 & 0.4998 &  27.659 &  10.042 \\
  0.014742 &  31.785640 & 0.1033 &  21.601 &  21.481 \\
  0.016278 &   8.736973 & 0.4522 &  28.300 &  13.052 \\
  0.018126 &   7.216778 & 0.4409 &  21.166 &  10.407 \\
  0.025971 &  $-$2.166898 & 0.5051 &  26.796 &   9.535 \\
  0.027360 &  21.655266 & 0.5380 &  26.743 &   8.559 \\
 \hline
 \end{tabular}
\end{table}

In Table~\ref{tab:catalogue} we provide a sample of the CAMIRA SDSS
DR8 cluster catalogue. The full version of the table will be available
in the online edition of the journal. 

\label{lastpage}

\end{document}